\documentclass[twocolumn,nofootinbib,superscriptaddress]{revtex4-1}
\usepackage{latexsym,epsfig,amssymb, amsmath,nicefrac}

\usepackage{color}

\usepackage[makeroom]{cancel}
  \usepackage{latexsym}
  \usepackage{epsf}
  \usepackage{amssymb}
  \usepackage{graphicx}
  \usepackage{amsmath}
  \usepackage{amsmath,amssymb,amsthm}
  \usepackage{verbatim}
  \usepackage{hyperref}

\def\bi{\begin{itemize}}
\def\ei{\end{itemize}}
\def\be{\begin{equation}}
\def\ee{\end{equation}}
\newcommand{\bea}{\begin{eqnarray}}
\newcommand{\eea}{\end{eqnarray}}

\def\lsim{\mathrel{\mathop
  {\hbox{\lower0.5ex\hbox{$\sim$}\kern-0.8em\lower-0.7ex\hbox{$<$}}}}}
\def\gsim{\mathrel{\mathop
  {\hbox{\lower0.5ex\hbox{$\sim$}\kern-0.8em\lower-0.7ex\hbox{$>$}}}}}

\begin{document}

\vspace{1cm}
\title{Effects of the speed of sound at large-$N$}

\author{Ivonne Zavala}

\email{e.i.zavalacarrasco@swansea.ac.uk}

\affiliation{Department of Physics, Swansea University, Singleton Park, Swansea, SA2 8PP, UK }

\begin{abstract}
The implications of a smaller than unity  speed of sound of the scalar perturbations during cosmic  inflation is studied in terms of a model independent large-$N$ approach. We show that the $(n_s, r)$ plane  is non-trivially degenerated when the effects of a different than one speed of sound $c_s$  are taken into account. We also discuss the  degeneracies due to $c_s$ in the running of the spectral index, $\alpha_s$. We use  the bounds on $c_s$  to constrain such degeneracies. 

\end{abstract}	

\maketitle

\section{Introduction}

In order to understand the large scale homogeneity and isotropy of the observable universe today, a period of accelerated expansion in the early universe, or cosmic inflation, is needed  \cite{Guth,Linde}. Moreover,  inflation also provides the seeds of the quantum fluctuations, which gave rise to the large scale structures in the universe that we observe today \cite{MCh}. 
Presently, hundreds of different models of inflation driven by a single scalar field exist,  with a wide variety of values for the parameters describing the amplitude of the spectra of quantum  fluctuations \cite{EI}.  

The stage of inflation and its predictions for the quantum fluctuations, can be  described without making precise reference to the microscopic mechanism generating the accelerated expansion  
\cite{Mukhanov} (see also \cite{Lotfi2,Lotfi1}). 
In particular,  inflation makes robust predictions for the primordial inhomogeneities. Namely, they are adiabatic,  approximately scale invariant and nearly Gaussian. Furthermore, inflation also predicts the existence of primordial gravitational waves.

All these properties are in perfect agreement with current observations. The latest reported values from the Planck collaboration for the  inflationary parameters are)\cite{Planck} (at $68\%$ {\rm c.l.})
\be\label{nsPlanck}
n_s -1 =  - 0.0397\pm0.0073\,,
\ee
for the spectral tilt, confirming  a percent-level deviation from a scale invariant spectrum, which corresponds to $n_s=1$. 
Furthermore, the scale dependence of the spectral index, or the running, defined  as $\alpha_s = d n_s/d\ln k$
has been constrained  to be
\be\label{alpha}
\alpha_s = -0.0134 \pm 0.009\,. 
\ee
In addition, the  Planck reported bound on primordial  equilateral  non-Gaussianity parameter at $68\,\%$c.l.~ is\footnote{Update: the new constraints in the cosmological parameters above reported by Planck15 \cite{Planck15Infla,Planck15NG}  are: $n_s =  0.968\pm0.006$; $\alpha_s=-0.003 \pm 0.007$; $f_{NL}^{\rm eq} = -16 \pm 70 $.}  $f_{NL}^{\rm eq} = -42 \pm 75 $ \cite{Planck2}. This type of non-Gaussianity results from models where the speed of the inflation perturbations can differ from unity. In this case, the latest Planck results report a lower bound for this speed of (95\% {\rm c.l.})
\be\label{csb}
c_s\geq 0.02  \,,\,\,
\ee
in an effective field theory parametrisation \cite{Cheung,Weinberg}. 

Further, the  observational upper bound for the ratio of the tensor-to-scalar mode, $r$, reported by Planck $95\,\%$ {\rm c.l.}~is
\be
r<0.11,
\ee
while determination of the primordial origin of the recently reported larger value of $r\sim 0.2$ by the BICEP2 collaboration \cite{bicep} awaits assessment.

The experimental values for the spectral index  and $r$  above, can be neatly described, independently of the underlying  model of inflation, when written solely in terms of the number of  e-folds $N$ left to the end of inflation after horizon crossing. 
The number of e-folds is  defined as $a = a_f \exp(-N)$, with $a$  the scale factor encoding the accelerated expansion.  The relevant interval of $N$ for observations is around $N_\star\sim 50-60$. For these values, the current data suggest a  $1/N_\star$ departure from a scale invariant spectrum \eqref{nsPlanck}.  

We consider the experimental data on $n_s$ and $r$ as an indication  to follow  a large-$N$ approach to inflation   \cite{Mukhanov,Diederik,GBR,GRSZ1,GRSZ2,Creminelli}. In particular, we  take carefully into account the possible degeneracies in the spectral index  and tensor-to-scalar ratio due to a smaller than unity speed of sound $c_s$. 
A smaller than unity speed of sound for the density perturbations during the inflationary stage, induces primordial non-Gaussianities \cite{AST} of equilateral shape and therefore it is constrained by data. Without specifying a concrete model for cosmic inflation, nevertheless, we cannot at present exclude a smaller than light speed for the perturbations.

\section{Speed of sound degeneracies  at large-$N$} 

The  generic formula for the spectral index $n_s$ in terms of the  slow variation parameters, defined as $\epsilon_1 \equiv d\ln H/ dN$, $\epsilon_2\equiv  d\ln \epsilon_1 /dN$ and $s\equiv d\ln c_s /dN$, is given by \cite{GM}
\be\label{ns}
n_s-1 = -2\,\epsilon_1 + \epsilon_2 + s  \,,
\ee
where  $c_s$ is the speed of sound, which can be different from unity in general models of cosmic inflation, giving rise to primordial non-Gaussianity of the equilateral shape  \cite{AST}.  The  running of the spectral index is further found from its definition, as $\alpha_s = \frac{1}{(\epsilon_1-1)} \frac{dn_s}{dN} $.
Furthermore, the tensor-to-scalar ratio in the more general case is given by \cite{GM}
\be\label{r}
r = 16\,c_s\, \epsilon_1\,.
\ee

The current experimental value for the spectral tilt suggests that it should scale as an inverse power of the number of e-folds $N$  at least to leading order in a large-$N$ expansion, independent of the source driving  inflation. That is, 
\be\label{nsN}
n_s -1 = -\frac{A}{N} + \dots
\ee
with $A$ some constant of order one and the dots referring to subleading terms in a $1/N$ expansion. 
From \eqref{nsN} and \eqref{ns}, we immediately see that the source of the $1/N$ behaviour of the spectral tilt can originate from any of the slow variation parameters.  One can simply assume that only one of them is responsible for such scaling,  while the others are subleading. For a unity speed of sound, one can identify two classes of models, according to whether $\epsilon_1$ and $\epsilon_2$ scale both as $1/N$ or only $\epsilon_2$ does \cite{Diederik}. 
In the present case, one can make a similar classification, including the speed of sound parameter $s$.
Therefore we consider the following expansions\footnote{One could in principle add a third class of models by asking that only $s\sim 1/N$, while $\epsilon_i$ are subleading in a large-$N$ expansion. It is not clear if successful models of inflation can have such behaviour, therefore we do not explore this case further.} 
\bea\label{largeNexp}
 \epsilon_1 = \frac{\alpha}{N^p} \quad \Rightarrow \quad \epsilon_2 = -\frac{p}{N}    \,,\qquad  \quad s = -\frac{\beta}{N}  \,,
\eea
where $\alpha>0$ and $\beta$ are some constants.  This implies that the speed of sound scales with $N$ as
\be\label{cs}
   c_s = \frac{c_0}{N^\beta}\,,
\ee
where $c_0$ is a positive integration constant, not necessarily of order unity\footnote{We are considering that scalar and tensor modes cross the horizon at similar instants \cite{GM}. One can check that modifications due to this effect \cite{AB} are negligible for our discussion. Constraints on the speed of sound taking into account this effect have been studied in \cite{BGP,PS}.}. 
Plugging \eqref{largeNexp} into \eqref{ns}-\eqref{csb} we  arrive at 
\be\label{nsr2}
n_s -1 = -\frac{2\alpha}{N^p} - \frac{p}{N} -\frac{\beta}{N},  \qquad \quad r= \frac{16\,c_0\,\alpha}{N^{\beta + p}} \,.
\ee
 Current observations indicate that $A\sim 2$ in  \eqref{nsN}, which restricts the values of  $\alpha, \beta$.
 
We now want to assess the degeneracies in the observational parameters ($n_s, r$), and the effects on the spectral tit running as we allow for a different than unity speed of sound. We do this by changing the values of $\beta$, taking into account the experimental bounds on $c_s$.  

\begin{figure}[htb]
\hspace{-3mm}
\begin{center}
\includegraphics[width=8.5cm]{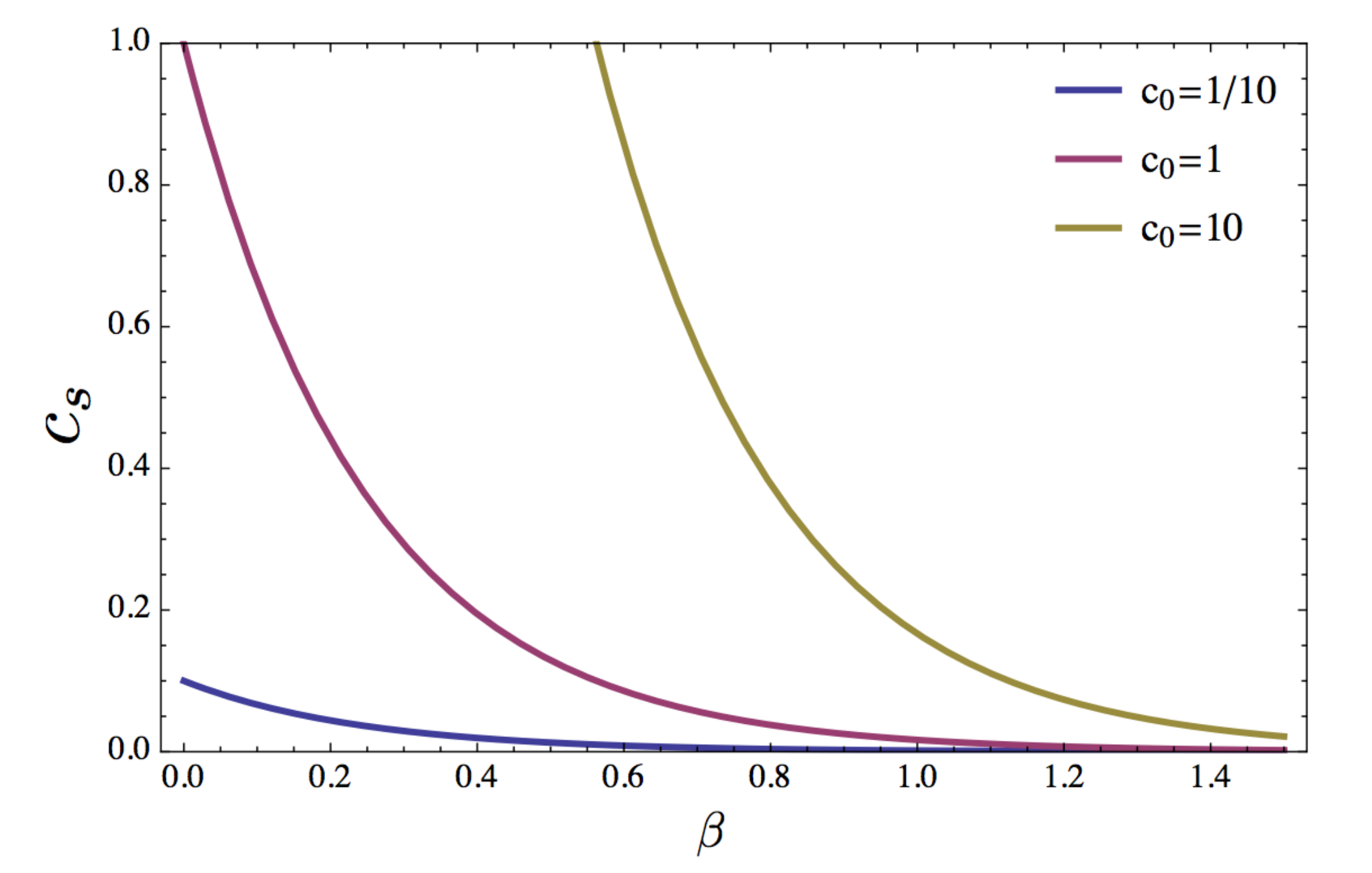}
\caption{Behaviour of the speed of sound as a function of $\beta$ for different values of $c_0$. }\label{fig2}
\end{center}
\end{figure}

Using the current bounds on $c_s$, we see that for a given value of the constant $c_0$, there is a maximum value for the $\beta$ parameter given by
\be\label{bmax}
\beta_{max} = \frac{\ln c_0}{\ln N} - \frac{\ln 0.02}{\ln N},  
\ee
where $N$ is evaluated at horizon crossing, which we take to be $N_\star=60$ for concreteness, but other values can be studied in a similar fashion. Further, the requirement that $c_s\leq 1$, sets a minimum value for $\beta$, given $c_0$:
\be\label{bmin}
\beta_{min} = \frac{\ln c_0}{\ln N}\,. 
\ee
Therefore, for each value of $c_0$, there is a range of values for $\beta$  within the observable window of $c_s$.  We illustrate this in figures  \ref{fig2}, where we  plot  $c_s$ as function of $\beta$ for different values of $c_0$. The allowed range of $\beta$ set by $c_s$   is evident from  this  figure. 
\begin{figure}[htb]
\hspace{-3mm}
\begin{center}
\includegraphics[width=8.7cm]{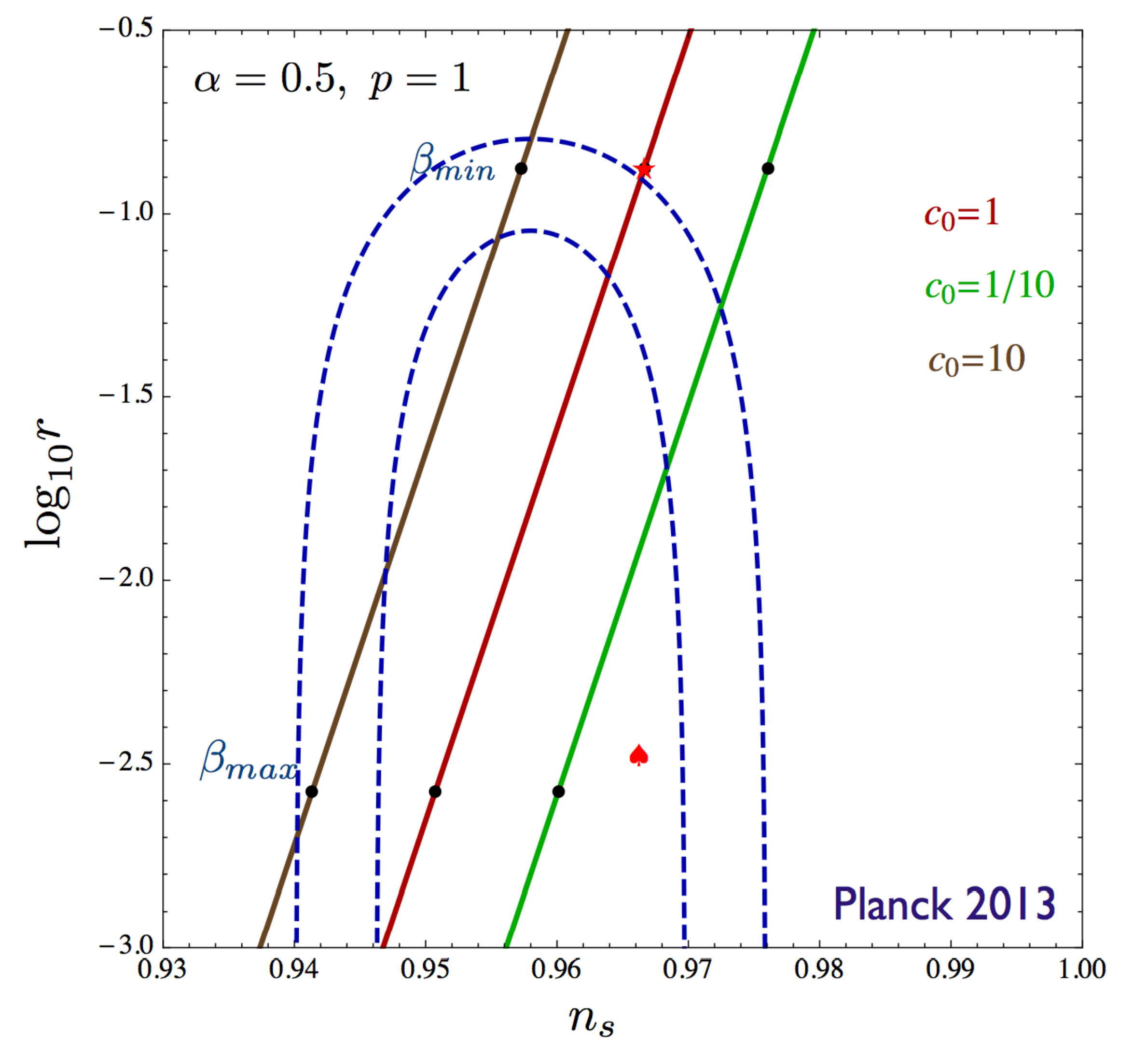}
\caption{Speed of sound effects in the $(n_s, {\rm log}_{10} r)$ plane for different values of $c_0$ for $\alpha=0.5$ and $p=1$. The dashed curves correspond to the experimental  Planck regions ($1$ and $2\sigma$ contours). The dark dots correspond to the allowed range of  $\beta$ set by the bounds on $c_s$, for the corresponding $c_0$. For example, for $c_0=1$, $(\beta_{min},\beta_{max})= (0, 0.96)$. The red marks correspond to the points of quadratic chaotic (star) and Starobinsky inflation (spades). This plot illustrates the shifts in $n_s, r$ by turning on the speed of sound, populating the intermediate regions between known slow roll models of inflation. }\label{fig3}
\end{center}
\end{figure}

\begin{figure}[htb]
\hspace{-3mm}
\begin{center}
\includegraphics[width=8.5cm]{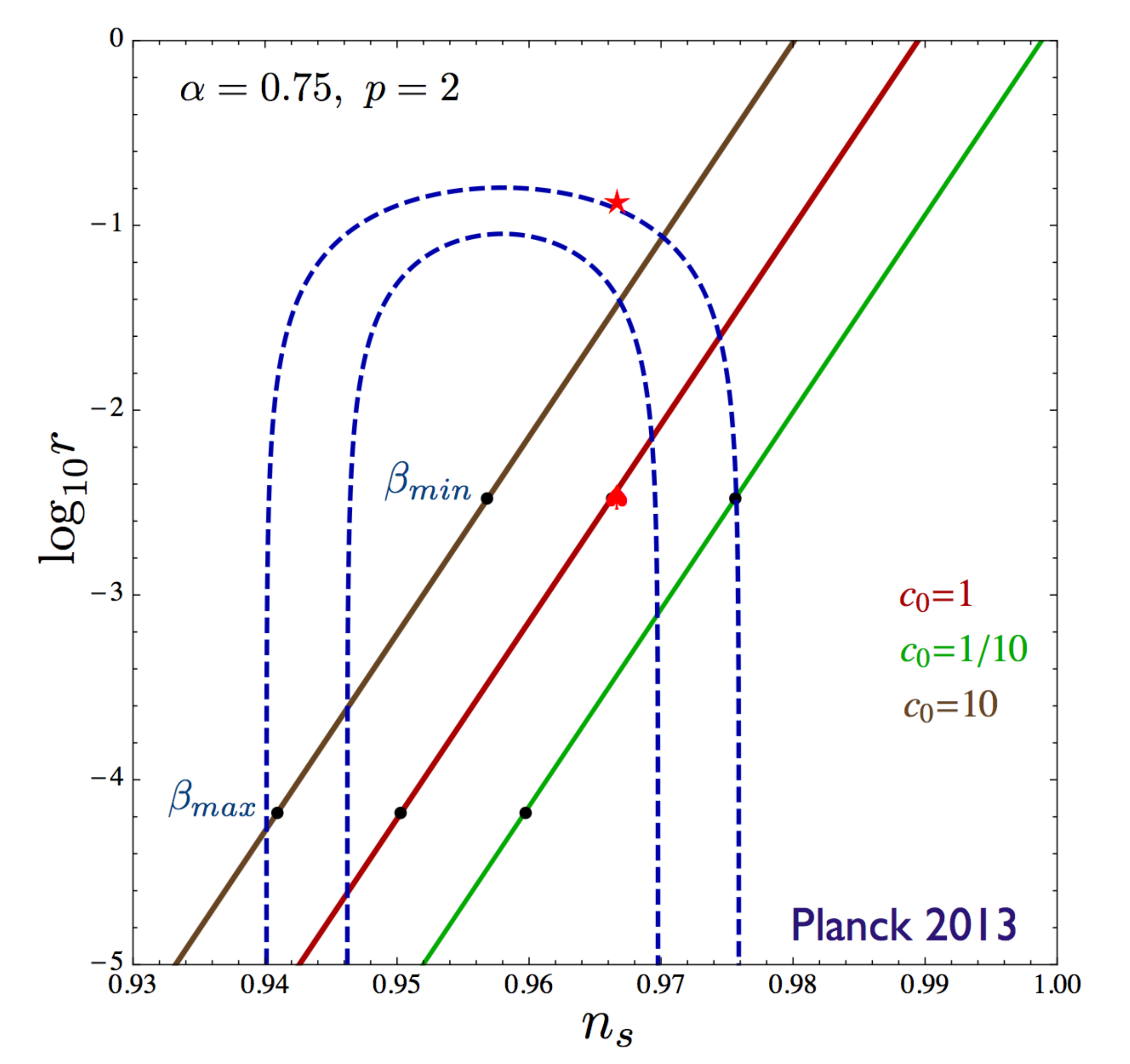}
\caption{Similar plot as in  figure \ref{fig3} of the speed of sound effects  in the $(n_s, {\rm log}_{10} r)$ plane, this time  for $\alpha=0.75$, $p=2$. 
}\label{fig4}
\end{center}
\end{figure}

The degeneracy sourced by the speed of sound in the spectral tilt and the tensor-to-scalar ratio is illustrated in figure \ref{fig3} for $\alpha =0.5$, $p=1$ and in figure \ref{fig4} for $\alpha=0.75$, $p=2$. In these pictures we see the displacement  along $c_s$, which is reflected in the change of $\beta$, for different values of $c_0$.  The  dots correspond to the  $(\beta_{min}, \beta_{max})$ range for the given value of $c_0$, corresponding  to the bounds on $c_s=(1,0.05)$. As a reference, the  chaotic and Starobinsky inflation points are marked with a star and a spade, respectively. These correspond to $\beta_{min}=0$  ($c_s=1$), $c_0=1$ in the figures. 
As we can see, the nontrivial effect of  the speed of sound, (encoded in  $\beta$ and $c_0$) is to  populate regions in the parameter space $(n_s, r)$ uncovered by the $c_s=1$ case. 

\begin{figure}[htb]
\hspace{-3mm}
\begin{center}
\includegraphics[width=8.3cm]{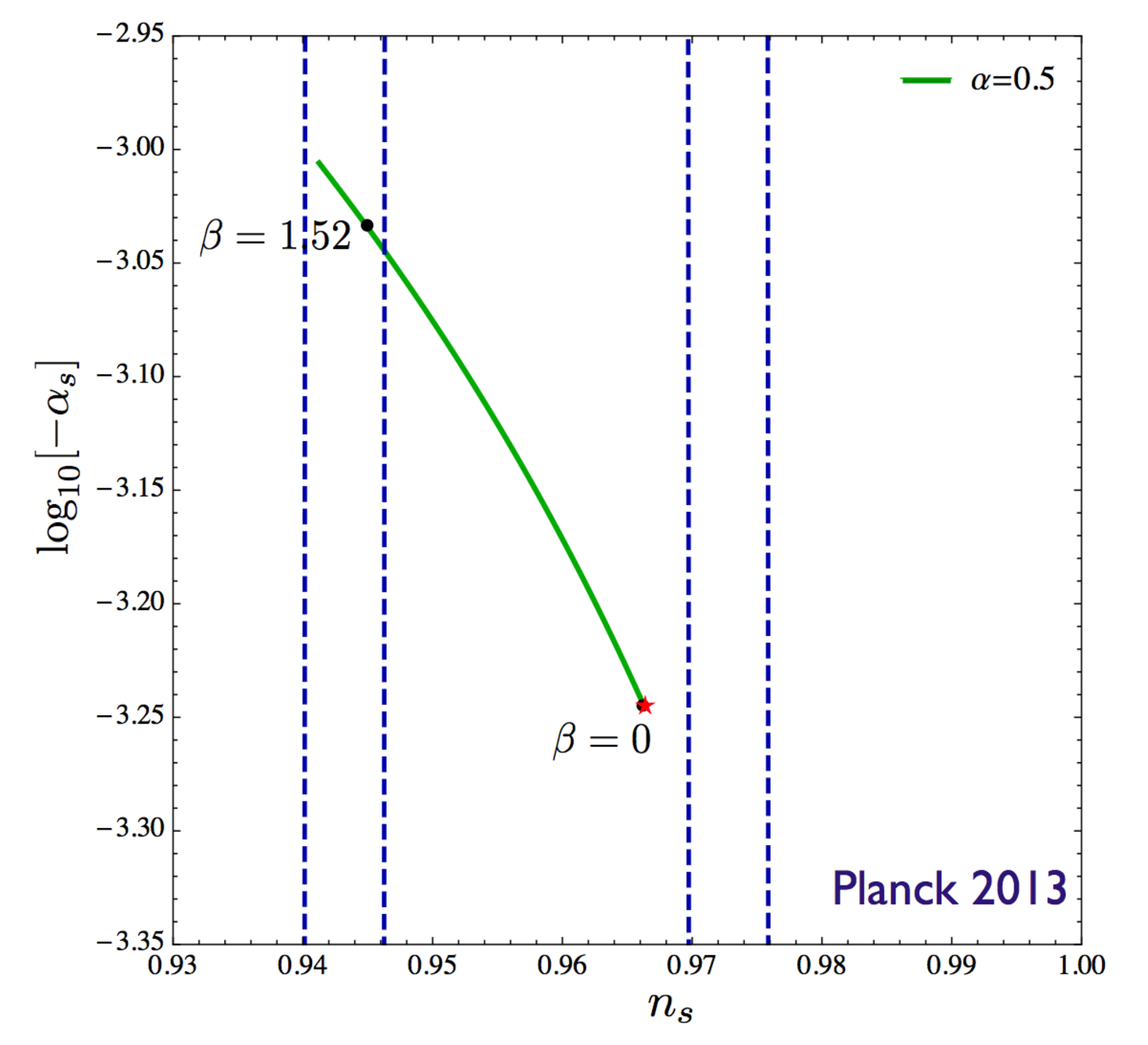}
\caption{The speed of sound effects in the running of the spectral index for $\alpha=0.5$. The dashed curves correspond to the allowed Planck regions. The  dots correspond to the indicated values of $\beta$, with $\beta_{max}=1.52$ corresponding to $c_0=10$ . As it is clear form \eqref{nsr2}, the effect of the speed of sound is to increase by a small amount the value of $\alpha_s$ as we increase $\beta$.  
}\label{fig5}
\end{center}
\end{figure}

Finally, we show the effect of the speed of sound for the running of the spectral index $\alpha_s$ in figure \ref{fig5} for $\alpha=0.5$. As it is evident from \eqref{nsr2}, variations in $c_s$ move the points in the $(n_s, \alpha_s)$ plane left upwards. Note however that the shifting is rather mild. From the large-$N$ expression for $\alpha_s$ we see that a similar shifting  occurs when varying  $\alpha$ and/or $p$.

As stressed above, we have taken the  $1/N$ behaviour of the spectral tilt  suggested  by observations, without prejudice about  the microscopic origin of the stage of inflation and the produced quantum fluctuations. 
In doing this, one should consider the implications of a different than unity speed of sound for the observable parameters, in particular $n_s$, $ r$ and $\alpha_s$. Moreover, the latest observational data  allow us to use the non-Gaussianity parameters to constrain this variable, as we have discussed. 

The important conclusion to draw from this analysis is that, if we do not focus on a particular model for producing the cosmological perturbations, models where some mechanism allows for a non-trivial speed of sound are clearly not excluded by current observations. Moreover, a different than unity  speed of sound pushes the tensor-to-scalar ratio and the spectral index to smaller values, covering regions in the ($n_s, r$) plane, which would seem empty for models with $c_s=1$. 
 On the other hand, it increases the running of the spectral tilt by only  a small amount.

\section{Discussion}

The current cosmological data indicate a suggestive $1/N$ departure from scale invariance in the spectral tilt when evaluated at the relevant range values of $N=50-60$. Remaining open about the precise origin of the inflationary stage and focusing in a more general description of the parameters in terms of a $1/N$ expansion,  one is  compelled to include all the small variation parameters contributing to the spectral tilt, \eqref{ns}. In particular, the speed of sound and its variation with $N$  give non-negligible contributions, with nontrivial implications  as we have discussed in this note. Variations in the speed of sound source non-Gaussianties, which allow us to constrain the variation on the speed of sound and the degeneracies implied in the $(n_s, r)$ plane and $\alpha_s$. 

We have not been concerned about constraining specific models or classes of models in our analysis. However, there are some single  scalar field models of inflation which may be relevant for our discussion. These are k- and DBI-inflation  \cite{ADM,ST,AST}, where $c_s$ can become smaller than unity.  In consistent single field  models of DBI inflation \cite{AZ}, it could be possible to allow for broad ranges of values for $\beta$ and therefore spectral tilt and tensor-to-scalar ratio, covering a large region in the $(n_s, r)$ plane, unpopulated by $c_s=1$ models. 

In the case of k- and DBI models of inflation, one may also consider the implications for the Lyth bound \cite{Lyth}. 
In the DBI  case, the Lyth bound  takes the same form as in the unity speed of sound case 
\be
\frac{\Delta\phi}{M_{Pl}} = \int{\sqrt{\frac{r}{8} }\,dN}\,,
\ee
as show in \cite{BM} (here $M_{Pl}$ is the Planck mass). However, since the speed of sound now enters into the tensor-to-scalar ratio (see \eqref{r}), the associated field range will be smaller as compared to the unity speed of sound case. One could nevertheless study the implications of $c_s$ for the field range along the lines of \cite{GRSZ1,GRSZ2} in a large-$N$ approach.

In summary, we have analysed the  implications of including the speed of sound in the general formulae for the observable inflationary parameters, namely the spectral tilt, and its running, as well as the tensor-to scalar ratio, independent of the mechanism producing the inflationary stage in the early universe. We have done so by writing all the slow variation parameters in  a $1/N$ expansion in the number of e-folds, as suggested by current observations. 
We have used the current constraints on the relevant non-Gaussianity parameter sourced by $c_s$ to explore and determine the implied degeneracies in the observable parameters. As we have seen, non-trivial motion in the $(n_s, r)$ plane can occur towards smaller values of  $n_s$ and $r$, covering parts of the parameter space, unpopulated by the $c_s=1$ case.

A possible way to break the degeneracies implied by the speed of sound, could come in the future from a scale dependence of the non-Gaussianity parameter, if non-Gaussianity is detected \cite{Chen,Gian}. 

It would also  be interesting to perform a model independent  analysis as has been done in \cite{Lotfi1,Lotfi2} taking into account the small variation parameter $s$.

\bigskip

\section*{Acknowledgments}
I would like to  thank Diederik Roest, Marco Scalisi and Gianmassimo Tasinato for useful comments on the manuscript.


\providecommand{\href}[2]{#2}\begingroup\raggedright\endgroup

\end{document}